\documentclass[
reprint,
longbibliography,
aps,
pra,
twoside,
superscriptaddress,
bibnotes,
floatfix
]{revtex4-2}
\usepackage{graphicx} 
\usepackage{physics}
\usepackage{dsfont}
\usepackage{color,newfloat}
\usepackage{graphicx}
\usepackage{dcolumn}
\usepackage{bm}
\usepackage{amsmath,amssymb,amsthm,amsfonts}
\usepackage{mathtools}
\usepackage{multirow}
\usepackage{booktabs}
\usepackage{array}
\usepackage{hyperref}
\usepackage[all]{hypcap}
\usepackage{chngcntr}
\usepackage{enumerate}
\usepackage[none]{hyphenat}
\usepackage[euler]{textgreek}
\usepackage{xcolor}
\usepackage{tabularx}
\usepackage{makecell}
\usepackage{booktabs}

\usepackage{subcaption}
\begin{document}
\title{Demonstration of a logical Bell-state measurement beyond the linear-optical limit}
\author{Shreya Kumar}
\affiliation{Institute for Quantum Information and Technologies, University of Stuttgart, 70569 Stuttgart, Germany}
\affiliation{Centre for Integrated Quantum Science and Technology (IQST), University of Stuttgart, 70569 Stuttgart, Germany}
\author{Simon D. Reiß}
\affiliation{Johannes-Gutenberg University of Mainz, Institute of Physics, Staudingerweg 7, 55128 Mainz, Germany}
\author{Peter van Loock}
\affiliation{Johannes-Gutenberg University of Mainz, Institute of Physics, Staudingerweg 7, 55128 Mainz, Germany}
\author{Stefanie Barz}
\email{barz@qit.uni-stuttgart.de}
\affiliation{Institute for Quantum Information and Technologies, University of Stuttgart, 70569 Stuttgart, Germany}
\affiliation{Centre for Integrated Quantum Science and Technology (IQST), University of Stuttgart, 70569 Stuttgart, Germany}

\begin{abstract}
Fault tolerance is essential for scalable quantum technologies and is enabled by quantum error-correction codes. Bell-state measurements (BSMs) are a fundamental building block for modern quantum technologies such as measurement-based quantum computation and fusion-based quantum computation, as well as quantum networks. Therefore, performing BSMs on error-corrected qubits is a necessary step for achieving fault tolerance in these applications. In this work, we realise a logical BSM using linear optics, based on a two-qubit repetition code, an instance of a quantum parity code that allows detection of bit-flip errors, and experimentally achieve a mean success probability of $(70.8\pm0.4)\%$. While standard linear-optical BSMs are fundamentally limited to a maximum success probability of $50\%$, this increased success probability enables higher secure key rates in quantum communication and facilitates the generation of large graph states for quantum computation. Since fault-tolerant schemes require error-correction codes regardless, this improvement comes at no additional resource overhead. Our results demonstrate that error-correction codes can be used to surpass the linear-optics limit of BSMs, which is an important step towards practical, fault-tolerant, and scalable photonic quantum technologies.
\end{abstract}
\maketitle
\begingroup
  \renewcommand{\thefootnote}{\fnsymbol{footnote}}
  \setcounter{footnote}{1}
  \footnotetext{These authors contributed equally.}
\endgroup

\section{Introduction}
Quantum error correction (QEC) is essential for realising fault-tolerant quantum information processing, as it enables the protection of quantum states against errors arising from imperfect physical systems~\cite{Knill1997Theory,bell2014experimental,gottesman1997stabilizer,PhysRevLett.98.190504}. In photonic architectures, photon loss is a particularly challenging error mechanism that must be addressed to enable scalable quantum technologies~\cite{wang2018,wasilewski2007protecting}. The quantum parity code (QPC), a family of codes that generalises the nine-qubit Shor code, provides a promising approach by redundantly encoding logical qubits across multiple photons, thereby offering intrinsic resilience against photon loss and enabling enhanced performance of logical operations, including logical Bell-state measurements~\cite{PhysRevA.52.R2493,PhysRevLett.95.100501,PhysRevLett.117.210501,PhysRevA.95.012327,PhysRevA.99.062308,PhysRevA.100.052303,reiss2026optimallogicalbellmeasurements}.

Bell-state measurements (BSMs) are a central primitive in photonic quantum information processing. They enable entanglement swapping in quantum repeaters and measurement-device-independent communication, provide the fusion operations required for generating large cluster states for measurement-based quantum computation (MBQC), and are an essential building block for fusion-based quantum computation (FBQC) and fault-tolerant quantum information processing~\cite{de2004,liu2013,tang2014,PhysRevLett.86.5188,PhysRevA.68.022312,PhysRevLett.95.010501,Lobl2025transforminggraph,Bartolucci2023}.

In photonic platforms, BSMs are commonly realised using linear optics due to their ease of implementation~\cite{weinfurter1994, pan1998,knill2001}. However, even in the absence of experimental imperfections, BSMs on qubits encoded in a single degree of freedom of a photon cannot be implemented with a $100\%$ success probability using passive linear optics alone~\cite{PhysRevA.59.3295,PhysRevA.69.012302}. Furthermore, without feedforward and ancillary photons, the success probability for discriminating uniformly distributed Bell states is fundamentally limited to 50\%~\cite{calsamiglia2001}.

This limitation can be addressed in several ways within the framework of linear optics. One approach is to employ external ancillary photons~\cite{ewert2014,PhysRevA.84.042331}, which has recently been demonstrated experimentally~\cite{bayerbach2023, hauser2025, guo2026}. Another approach is based on quantum error-correcting codes, in which the logical information is distributed across multiple physical qubits, allowing several physical BSMs to be combined into a logical BSM with a higher success probability. Therefore, in addition to protecting encoded information against physical errors, such encodings can be used to implement logical BSMs on encoded qubits, thereby enabling BSMs with success probabilities that exceed the linear-optics limit. Considerable theoretical work has focused on logical BSMs on stabiliser codes based on linear-optics BSMs~\cite{gottesman1997stabilizer,PhysRevLett.114.113603,PhysRevA.92.052324,PhysRevLett.117.210501,PhysRevA.95.012327,PhysRevA.99.062308,PhysRevA.100.052303,Hilaire_2021,PRXQuantum.4.040322,patil2024improveddesignallphotonicquantum,reiss2026optimallogicalbellmeasurements}. For photonic quantum computation, they can protect logical qubits against photon loss and enhance the success probability of Bell-state and fusion measurements required for MBQC and FBQC, thereby reducing resource overheads and improving the scalability of such architectures that rely on these probabilistic operations~\cite{Bartolucci2023,hauser2025}. For all-optical quantum repeaters, these can protect quantum information against errors such as loss and improve the scalability and efficiency of long-distance quantum communication~\cite{Azuma2015,PhysRevLett.114.113603,PhysRevA.92.052324,PhysRevA.100.052303,PhysRevLett.117.210501,PhysRevA.95.012327,Hilaire_2021,patil2024improveddesignallphotonicquantum,PRXQuantum.4.040322}. Despite these advances, an experimental implementation of logical BSMs has not yet been demonstrated.

In this work, we demonstrate a logical BSM using static linear optics on two logical qubits, each encoded in a two-qubit repetition code. As an instance of the QPC, the two-qubit repetition code represents a first step toward larger QEC codes and can detect a single-bit-flip error. This code enables a logical BSM that surpasses the fundamental $50\%$ limit of linear-optics BSMs for dual-rail-encoded qubits, achieving a theoretical success probability of $75\%$ in an ideal setting.
\begin{figure*}[ht!]
    \centering
    \includegraphics[width=1\linewidth]{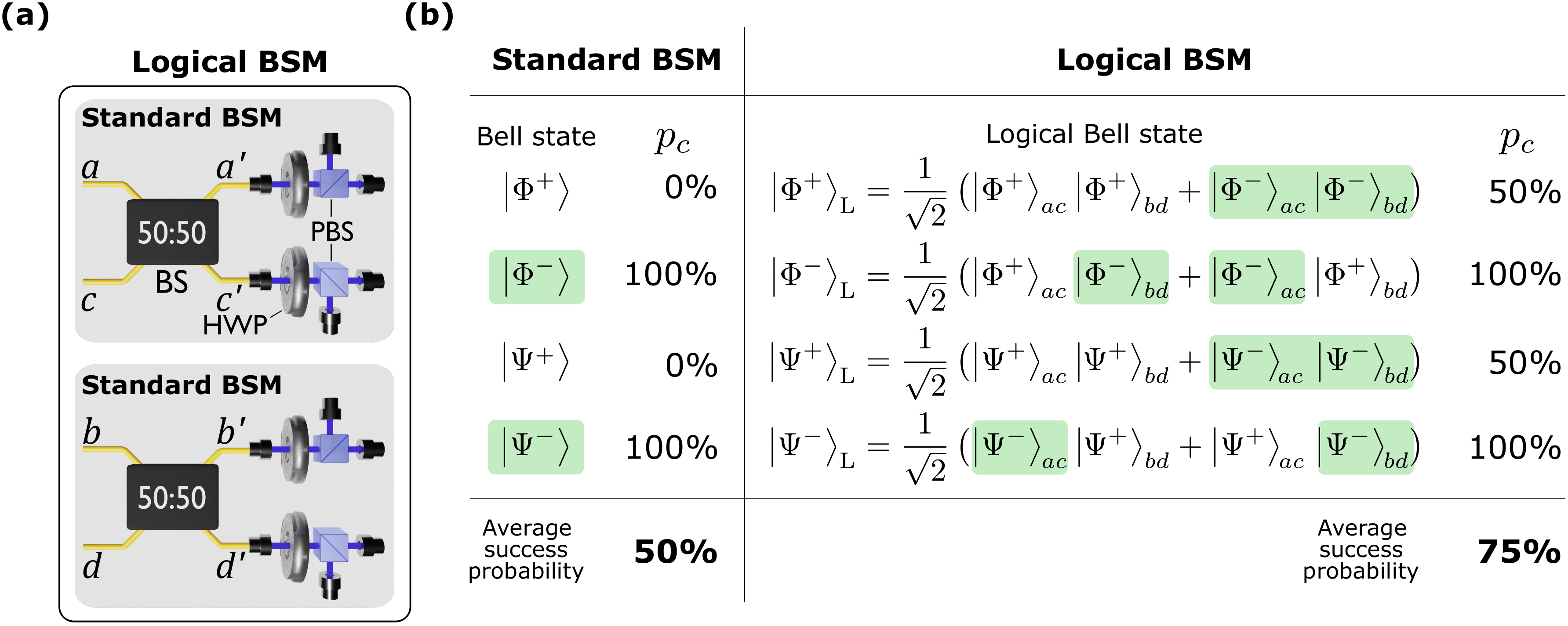}
    \caption{\textbf{Comparison between standard and logical Bell-state measurements (BSMs).} (a) Schematic representation of the logical BSM. Four physical qubits encode a logical Bell state. The logical BSM is implemented using two standard BSMs on the qubit pairs $(a,c)$ and $(b,d)$. The output modes are rotated by applying a Hadamard gate, and then polarisation-resolved using polarising beam splitters (PBSs). (b) Success probabilities associated with each of the four Bell states for the standard BSM (left) and the logical BSM (right). In the variant of a standard BSM that we are using in our experiment, the states $\ket*{\Phi^-}$ and $\ket*{\Psi^-}$ can be unambiguously identified, each with a success probability of $100$\%, whereas the states $\ket*{\Phi^+}$ and $\ket*{\Psi^+}$ lead to ambiguous outcomes. The logical encoding decomposes into the physical Bell states such that the logical Bell states \(\ket*{\Phi^+}_{\text{L}}\) and \(\ket*{\Psi^+}_{\text{L}}\) can be discriminated in $50$\% of the cases. The physical Bell states that lead to unambiguous results are highlighted with green boxes. The standard BSM has an average success probability of $50\%$, while the logical BSM increases this value to $75\%$. }
    \label{fig:Concept}
\end{figure*}

\section{Theory}
The QPC encodes quantum information in a way that is particularly resilient to photon loss. In QPC($n,m$), one logical qubit is encoded into $n$ blocks, each containing $m$ physical qubits. The computational basis states of a block are defined as
\begin{equation}
    \begin{aligned}
        \ket{0}^{(m)} = \ket{0}^{\otimes m}, \quad
        \ket{1}^{(m)} = \ket{1}^{\otimes m}.
    \end{aligned}
\end{equation}
and the logical basis states are defined in the logical $X_\text{L}$ basis as
\begin{equation}
    \begin{aligned}
        \ket{\pm}^{(n,m)}_{\text{L}} & = \frac{1}{\sqrt{2^n}} ( \ket{0}^{(m)} \pm \ket{1}^{(m)} )^{\otimes n}. \\
    \end{aligned}
\end{equation}
The code distance is $d=\min(n,m)$, which implies that it can correct up to $\lfloor(d-1)/2\rfloor$ arbitrary single-qubit errors. For photon loss, the error-correction capability is not determined solely by the code distance. The logical information can, in principle, be recovered, provided that no block is lost completely and at least one photon remains in every block~\cite{PhysRevLett.95.100501,PhysRevLett.117.210501,PhysRevA.95.012327}. For example, QPC($2,2$) has a code distance of $d=2$ and, therefore, cannot correct an arbitrary single-qubit error. However, we can recover the logical information after the loss of any single photon, since the encoded logical information is still fully contained in the remaining qubits. This enhanced protection against photon loss makes the QPC particularly attractive for both photonic quantum computing and communication.

Beyond its error-protection capabilities, the QPC also enables BSMs on encoded logical qubits with success probabilities that substantially exceed the $50\%$ limit of standard static linear-optical BSMs on dual-rail qubits. QPC($1,2$), equivalent to a two-qubit repetition code, enables a static linear-optical BSM with a success probability of $75\%$. Furthermore, it can detect a single-qubit flip error. Note that QPC($2,1$) is essentially identical to QPC($1,2$), up to a Hadamard rotation on the qubit basis, and can detect a single phase flip error.

We represent the logical computational basis states in the repetition encoding at the qubit level by:
\begin{equation}
    \begin{aligned}
        \ket{0}_{\text{L}} & = \ket{0}\otimes\ket{0},\\
        \ket{1}_{\text{L}} & = \ket{1}\otimes\ket{1},
    \end{aligned}
    \label{eq:repetition-code}
\end{equation}
where each logical qubit is composed of two physical qubits. This code already detects bit-flip errors through a measurement of the Pauli operator $Z \otimes Z$. Experimentally, the measurement outcome determines the parity of the two physical qubits, from which the corresponding $ZZ$ eigenvalue of $+1$ or $-1$ is inferred. Obtaining the eigenvalue $-1$ signals that a bit flip has occurred on one of the two physical qubits, although the measurement does not identify which qubit was affected. Thus, the measurement provides error detection but not error correction. An eigenvalue of $+1$ indicates that no detectable error has occurred. In this encoding, the Bell states
\begin{equation}
    \begin{aligned}
        \ket{\Phi^\pm} & = \frac{1}{\sqrt{2}} \left(\ket{00} _{ab}\pm \ket{11}_{ab}\right), \\
        \ket{\Psi^\pm} & = \frac{1}{\sqrt{2}} \left(\ket{01}_{ab} \pm \ket{10}_{ab}\right),
    \end{aligned}
\end{equation}
where $\ket{0}$ and $\ket{1}$ are the computational basis states of the qubits $a$ and $b$, are given by
\begin{equation}
    \begin{aligned}
        \ket{\Phi^\pm}_{\text{L}} & = \frac{1}{\sqrt{2}} \left(\ket{0000}_{abcd} \pm \ket{1111}_{abcd}\right), \\
        \ket{\Psi^\pm}_{\text{L}} & = \frac{1}{\sqrt{2}} \left(\ket{0011}_{abcd} \pm \ket{1100}_{abcd}\right),\\
    \end{aligned}
    \label{eq:logical_Bellstates}
\end{equation}
where the first logical qubit is encoded in the physical qubits $a$ and $b$, and the second logical qubit is encoded in the physical qubits $c$ and $d$.

We rearrange the order of the second and third modes and rewrite the logical Bell states in the physical Bell basis \cite{PhysRevLett.114.113603,PhysRevA.92.052324,PhysRevLett.117.210501}:
\begin{equation}
    \begin{split}
        \ket{\Phi^\pm}_{\text{L}} & = \frac{1}{\sqrt{2}} \left(\ket{\Phi^+}_{ac} \ket{\Phi^\pm}_{bd} + \ket{\Phi^-}_{ac} \ket{\Phi^\mp}_{bd}\right)\\
        \ket{\Psi^\pm}_{\text{L}} & = \frac{1}{\sqrt{2}} \left(\ket{\Psi^\pm}_{ac} \ket{\Psi^+}_{bd} + \ket{\Psi^\mp}_{ac} \ket{\Psi^-}_{bd}\right)
        \label{eq:Bell-states-encoded-rearranged}
    \end{split}
\end{equation}

We implement a logical BSM by performing one standard linear-optical BSM on each of the two qubit pairs $(a,c)$ and $(b,d)$.

A widely used physical implementation of photonic qubits is encoding them in the polarisation degree of freedom, with the computational basis states for qubit $a$ identified as $\ket{0} \equiv a^\dagger_0 \ket{\mathrm{vac}}=\ket{H}$ and $\ket{1} \equiv a^\dagger_1 \ket{\mathrm{vac}}=\ket{V}$, where $\ket{H}$ and $\ket{V}$ denote horizontal and vertical polarisation, respectively. Nevertheless, the following discussion can be equivalently adapted to other qubit encodings such as path encoding. In polarisation encoding, the Bell states can be written as:
\begin{equation}
    \begin{aligned}
        \ket{\Phi^\pm} & = \frac{1}{\sqrt{2}} \left( a_H^\dagger c_H^\dagger \pm a_V^\dagger c_V^\dagger \right) \ket{\mathrm{vac}}, \\
        \ket{\Psi^\pm} & = \frac{1}{\sqrt{2}} \left( a_H^\dagger c_V^\dagger \pm a_V^\dagger c_H^\dagger  \right) \ket{\mathrm{vac}}, 
        \label{eq:Bell-state-operator-notation}
    \end{aligned}
\end{equation}
where $a_i^\dagger$ and $c_i^\dagger$ are the creation operators for the qubits `$a$' and `$c$' in the polarisation $i \in \{H,V\}$. A linear-optical BSM can be implemented using a beam splitter followed by polarisation-resolved measurements on each output mode~\cite{PhysRevA.51.R1727}. Additionally, we apply a Hadamard gate to each of the output modes of the beam splitter using half-wave plates set to $22.5^\circ$. The Hadamard gates permute the correspondence between Bell states and output states, changing which Bell states correspond to which measurement outcomes. This altered correspondence is leveraged in our logical Bell-state measurement. Under the combined action of the beam splitter and the Hadamard gates, the Bell states transform as follows:
\begin{equation}
    \begin{aligned}
        \ket{\Phi^+} & \rightarrow \frac{1}{2\sqrt{2}} \left(a'^\dagger_H a'^\dagger_H + c'^\dagger_H c'^\dagger_H + a'^\dagger_V a'^\dagger_V + c'^\dagger_V c'^\dagger_V\right) \ket{\mathrm{vac}}, \\
        \ket{\Psi^+} & \rightarrow \frac{1}{2\sqrt{2}} \left(a'^\dagger_H a'^\dagger_H + c'^\dagger_H c'^\dagger_H - a'^\dagger_V a'^\dagger_V - c'^\dagger_V c'^\dagger_V\right) \ket{\mathrm{vac}}, \\
        \ket{\Phi^-} & \rightarrow \frac{i}{\sqrt{2}} \left(a'^\dagger_H a'^\dagger_V + c'^\dagger_H c'^\dagger_V\right) \ket{\mathrm{vac}}, \\
        \ket{\Psi^-} & \rightarrow \frac{i}{\sqrt{2}} \left(a'^\dagger_H c'^\dagger_V - a'^\dagger_V c'^\dagger_H\right) \ket{\mathrm{vac}}, 
    \end{aligned}
    \label{eq:transformed-state}
\end{equation}
where $a'^\dagger_{i}$, and $c'^\dagger_{i}$ ($i \in \{H,V\}$) are creation operators for the two output spatial modes $a'$ and $c'$ and polarisation mode $i$. Assuming no loss and on-off click detectors, we can unambiguously discriminate $\ket{\Phi^-}$ and $\ket{\Psi^-}$ from each other as well as against $\ket{\Psi^+}$ and $\ket{\Phi^+}$. However, if we detect two photons in one of the output modes, we cannot discriminate between $\ket{\Phi^+}$ and $\ket{\Psi^+}$. Since one can still distinguish between the sets $\{ \ket{\Phi^+}, \ket{\Psi^+}\}$ and $\{ \ket{\Phi^-}, \ket{\Psi^-}\}$ in this case, this is commonly referred to as a partial result. Therefore, assuming a uniform mixture of Bell states, we achieve a success probability of $50\%$.
\begin{figure*}[ht!]
   \centering
    \includegraphics[width=1\textwidth]{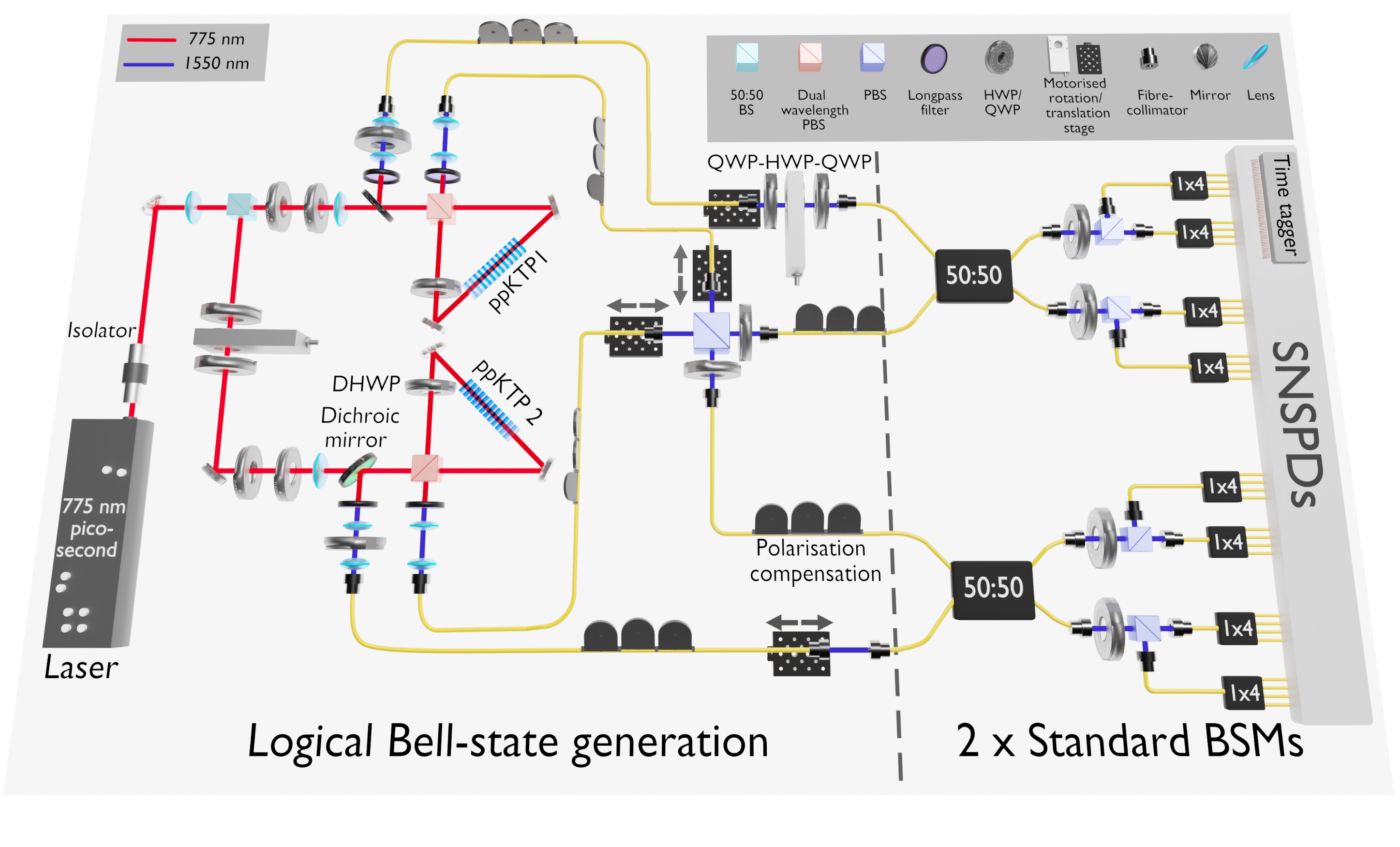}
    \caption{\textbf{Experimental implementation of the logical BSM.} The setup is primarily composed of two parts: The logical Bell-state generation and the logical BSM. \textbf{Logical Bell-state generation:} A picosecond pulsed $775$~nm laser pumps two spontaneous parametric down-conversion sources that are based on type-II ppKTP crystals to generate entangled photon pairs at $1550$~nm. One photon from each pair is made incident at the two ports of a PBS. Measuring four-fold coincidences results in the generation of a state that is subsequently transformed to one of the Bell states using wave plates. We use motorised translation stages to ensure that the photons are indistinguishable in their arrival time, and wave plates to switch between different logical Bell states. \textbf{Logical BSM:} Two standard BSMs realise one logical BSM. We implement this using two fibre-based beam splitters, followed by PBSs to resolve the polarisation modes. Each mode is demultiplexed into four detector channels, allowing for pseudo-photon-number resolution. }
    \label{fig:experimental_implementaion}
\end{figure*}

Using the decomposition in Eq.~\eqref{eq:Bell-states-encoded-rearranged}, we see that for the logical Bell states $\ket{\Phi^-}_{\text{L}}$ and $\ket{\Psi^-}_{\text{L}}$, at least one of the two standard BSMs on $(a,c)$ and $(b,d)$ is guaranteed to result in an unambiguous outcome of $\ket{\Phi^-}$ and $\ket{\Psi^-}$, respectively, as shown in Fig.~\ref{fig:Concept}. Therefore, both $\ket{\Phi^-}_{\text{L}}$ and $\ket{\Psi^-}_{\text{L}}$ are identified with a success probability of $100\%$. 

The logical Bell states $\ket{\Phi^+}_{\text{L}}$ and $\ket{\Psi^+}_{\text{L}}$, on the other hand, either lead to two unambiguous or two ambiguous outcomes across the two standard BSMs, each with equal probability (see Fig.~\ref{fig:Concept}(b)). The unambiguous outcomes result in additional detection (click) patterns that allow us to discriminate between $\ket{\Phi^+}_{\text{L}}$ and $\ket{\Psi^+}_{\text{L}}$, whereas the ambiguous outcomes correspond to failed measurements. In six of the eight cases, the two standard BSMs unambiguously identify the Bell states, leading to successful identification of the logical Bell state. Two of the cases result in an ambiguous result for the states $\ket{\Phi^+}_{\text{L}}$ and $\ket{\Psi^+}_{\text{L}}$, each with a 50\% probability. Averaging over the four Bell states results in an overall success probability of $75$\%:
\begin{equation}
p_c = \frac{1}{4}\sum_{i \in \{\Psi^{\pm},\Phi^{\pm}\}} p_{\ket{i}}=\frac{1}{4}\left(0.5+1.0+1.0+0.5\right)= 0.75.
    \label{eq:successprobability}
\end{equation}
Note that another approach to understanding the logical BSM is via the stabiliser formalism, which is presented in Appendix.~\ref{app:stabilisers}.

The success probability of a logical BSM on a repetition code already achieves the optimal scaling with the number of physical qubits, $m$, per logical qubit, reaching $1-2^{-m}$ using static linear optics~\cite{PhysRevLett.117.210501,PhysRevLett.114.113603,PhysRevA.92.052324,PhysRevA.100.052303}. Extending to a larger QPC$(n,m)$ would provide genuine error-correction capabilities. Two classes of linear-optical schemes can be distinguished: static schemes without feedforward, and adaptive schemes. Ref.~\cite{PhysRevA.99.062308} presented a static linear-optics scheme with a no-loss success probability of $1-2^{-(n+m-1)}$ for QPC($n,m$). By incorporating feedforward, the success probability can be further improved to $1-2^{-nm}$, as demonstrated in Ref.~\cite{PhysRevA.100.052303}. For a fixed total number of physical qubits, the code distance is maximised by choosing balanced codes with $n=m$. In this case, success probabilities of $99\%$ are achieved with 16 physical qubits for the static scheme and 9 physical qubits for the adaptive scheme.

\section{Experiment}

\begin{figure*}
    \centering
    \includegraphics[width=1\linewidth]{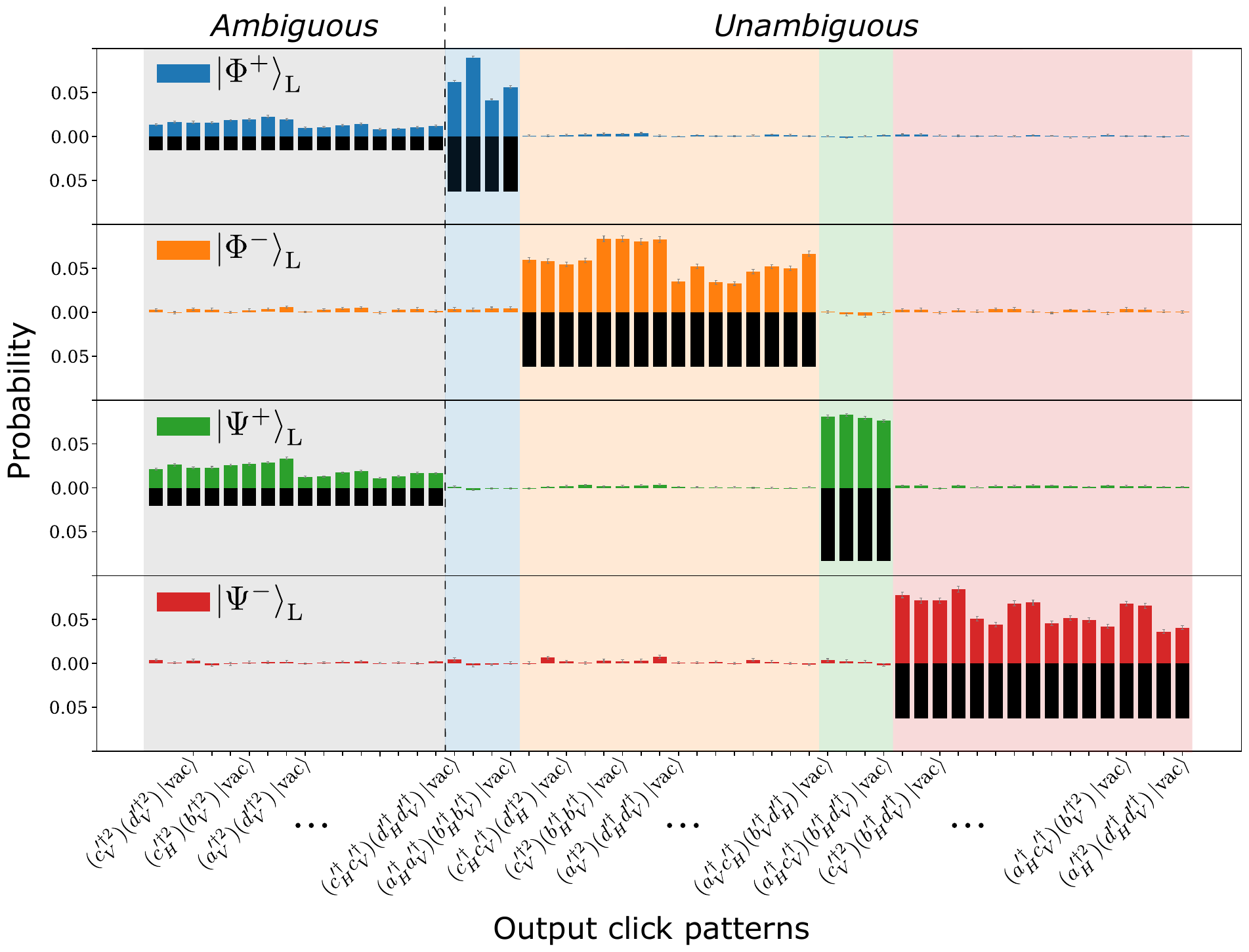}
    \caption{\textbf{Comparison of experimental and theoretical click patterns for the logical Bell-state measurement.} The four panels correspond to the click patterns observed for each of the four logical Bell states. The output modes of the two physical BSMs are grouped and indicated in brackets. The measured probability distribution is represented by the coloured bars, while the theoretical values are plotted as mirrored bars for comparison. 
    The encoded scheme allows for a subset of outcomes from the standard scheme to be correctly classified, successfully increasing the total average efficiency to $(70.8\pm0.4)\%$, which is beyond the $50$\% threshold. Background colours highlight the click patterns that lead to ambiguous results (grey), and detection click patterns that result in unambiguous results (coloured). The error bars are obtained from the variance of the measured data.}
    \label{fig:Results}
\end{figure*}

We implement the logical BSM with the experimental setup shown in Fig.~\ref{fig:experimental_implementaion}. We generate the logical Bell states by interfering two photon pairs in the state $\ket{\Psi^-}$ at a polarising beam splitter (PBS). The two constituent Bell states are generated using spontaneous parametric down-conversion in type-II periodically poled potassium titanyl phosphate (ppKTP) crystals operated in a Sagnac interferometer configuration. One photon from each entangled pair is directed to the two input ports of the PBS. By post-selecting on four-fold coincidences and subsequently applying appropriate single-qubit rotations, the resultant state is mapped onto the logical Bell state $\ket{\Phi^+}_{\text{L}}$, as defined in Eq.~\eqref{eq:logical_Bellstates}: 
\begin{equation}
    \ket{\psi}_{\text{out}}=\frac{1}{\sqrt{2}}\left(\ket{HHHH}+\ket{VVVV}\right)=\ket{\Phi^+}_{\text{L}}.
    \label{eq:state_preparation}
\end{equation}
See Appendix~\ref{GHZ_Appx} for a detailed characterisation of the resource state. The other logical Bell states from Eq.~\eqref{eq:logical_Bellstates} are obtained by further applying single-qubit rotations.

The logical BSM is realised by two standard physical BSMs. Each physical BSM uses a fibre-based balanced beam splitter, followed by a Hadamard operation and polarisation-resolved detection with PBSs. Each PBS output is demultiplexed into four detector channels, enabling pseudo-photon-number resolution. Background correction is performed to subtract higher-order contributions from a single source to the four-fold coincidences.

\section{Results and discussion}

The measured probabilities for measuring different click patterns for the four logical Bell states are presented in Fig.~\ref{fig:Results}. Compared to a standard BSM, which can unambiguously discriminate only two out of the four Bell states, we obtain additional distinct and unambiguous detector click patterns for the logical states $\ket*{\Phi^+}_{\text{L}}$ and $\ket*{\Psi^+}_{\text{L}}$ in the measurement events, as highlighted by the blue and green panels. We extract the success probability from the measured data by summing the probabilities of measuring correct and unambiguous click patterns for a given Bell state. This corresponds to the patterns for specific Bell states that are highlighted under the `unambiguous' section of the plot in Fig~\ref{fig:Results}. 

The success probabilities of the two logical states, $\ket{\Phi^-}_{\text{L}}$ and $\ket{\Psi^-}_{\text{L}}$, that were already unambiguous under the standard BSM remain close to ideal under the logical scheme, at $(93.0 \pm 0.8)$\% and $(93.9 \pm 0.8)$\%, respectively. Additionally, the two states that rely on the logical encoding to be resolved reach success probabilities of $(48.7\pm0.6)$\% and $(47.5 \pm 0.5)$\% for the states $\ket{\Phi^+}_{\text{L}}$ and $\ket{\Psi^+}_{\text{L}}$, respectively. This is close to their $50$\% theoretical contribution, confirming that the encoding complies with the prediction in Eq.~\eqref{eq:successprobability}. Averaged over all four states, this gives an overall success probability of $(70.8 \pm 0.4)\%$ (Table~\ref{tab:Metrics}).
\begin{table}[b!]
\centering

\begin{tabularx}{\linewidth}{lccc}

\toprule
\textbf{State} & $\bm{p}_c$& \textbf{MDF} & \textbf{Distance} \\
\toprule
$\ket{\Phi^+}_{\text{L}}$ & $(48.7\pm0.6)$\% & $(87.2\pm1.3)\%$ & $0.165 \pm 0.007$ \\
$\ket{\Phi^-}_{\text{L}}$ & $(93.0\pm0.8)\%$ & $(96.3\pm0.7)\%$ & $0.168 \pm 0.007$ \\
$\ket{\Psi^+}_{\text{L}}$ & $(47.5\pm0.5)\%$ & $(87.5\pm1.0)\%$ & $0.130 \pm 0.005$ \\
$\ket{\Psi^-}_{\text{L}}$ & $(93.9\pm0.8)\%$ & $(95.6\pm0.7)\%$ & $0.148 \pm 0.007$ \\
\bottomrule
\textbf{Average} & $(70.8\pm0.4)\%$ & $(91.7\pm0.5)\%$ & $0.153 \pm 0.005$ \\
\bottomrule
\end{tabularx}
\caption{\textbf{Performance metrics for the logical BSM.} Success probability, measurement discrimination fidelity, and variation distance for the four logical Bell states.}
\label{tab:Metrics}
\end{table}
Errors in the state preparation and imperfect interference could lead to unambiguous click patterns, yet correspond to a Bell state different from the input state, thereby causing a false identification. We extract the probability of false identification, $p_f$, from the measured click patterns, and define the measurement discrimination fidelity (MDF) as~\cite{wein2016}
\begin{equation}
    \text{MDF}=\frac{p_c}{p_c+p_f}.
\end{equation}
For our implementation, we obtain an average MDF of $(91.7\pm0.5)\%$ for the four logical Bell states. 

Additionally, to compare the implemented scheme with the theory, we compute the total variation distance, $D$, which is obtained by summing the absolute difference between the experimentally observed click patterns and the probabilities predicted from theory. We calculate it using~\cite{wang2019}:
\begin{equation}
    D=\sum_i\frac{\left|f_i-q_i\right|}{2},
\end{equation}
where $f_i$ and $q_i$ are the experimental and theoretical probabilities for the $i$th click pattern, respectively. A smaller distance indicates a high degree of overlap between the physical implementation and the theoretically expected click pattern, serving as a benchmark for the overall performance of the implemented scheme. We measure an average distance of $0.153\pm0.005$, indicating a good agreement between the experimental results and theory. The three metrics given above are listed for the four individual logical Bell states in Table~\ref{tab:Metrics}. 

\section{Conclusion}
In this work, we report the first experimental implementation of a logical BSM on two logical qubits encoded in a two-qubit repetition code based on single photons and linear optics. In our experiment, we achieved a success probability of \mbox{$(70.8\pm0.4)$\%}, which is well beyond the linear-optics limit of the standard BSM.

This result demonstrates that quantum error-correcting codes can enhance fundamental quantum information processing primitives beyond their role in protecting quantum information. In particular, error-correction codes enable logical BSMs to surpass the $50$\% limit and mitigate the intrinsic probabilistic nature of linear-optical BSMs. Importantly, if such encodings are already required for fault-tolerant quantum information processing, the enhanced BSM performance is obtained without additional ancillary photons, allowing the encoding resources to provide error protection and increased measurement success simultaneously. Conceptually, this work contributes to the step from fundamental quantum information processing primitives to encoded, fault-tolerant implementation building blocks of photonic quantum technologies. 

A next step towards fault-tolerant quantum technologies is the implementation of logical BSMs on logical qubits encoded in QEC codes with non-trivial code distance and full error-correcting capabilities. Beyond providing enhanced error protection, larger codes would simultaneously increase the logical BSM success probability. This dual benefit makes logical BSMs naturally suited for scalable fault-tolerant photonic architectures. Ultimately, such logical BSMs may serve as a key primitive for scalable fault-tolerant quantum computation and long-distance quantum networking.
\section{Acknowledgements}
We thank Nico Hauser, Joscha Heinze and Lukas Ruckle for helpful discussions, and Simone Evaldo D'Aurelio for helping with the implementation of the data analysis script.  
We acknowledge the support from the Federal Ministry of
Research, Technology and Space (BMFTR, projects SiSiQ: FKZ 13N14920, PhotonQ: FKZ 13N15758, QR.N: FKZ 16KIS2207, TD.QR: FKZ 16KISS026, QCyber: FKZ 16KIS2590K), and the Deutsche Forschungsgemeinschaft (DFG, German Research Foundation, 431314977/GRK2642, 516238647/SFB 1667, 563437379/SPP2514), the Carl Zeiss Foundation, and the Centre for Integrated Quantum Science and Technology (IQST).

\section{Author contribution}

P.L. and S.R. established the theoretical framework for this work. S.B. and S.K. conceptualised the project. S.K. carried out the experimental work and curated the data. S.B. supervised the project. S.K. and S.R. drafted the manuscript, and all authors participated in its review and revision.

\section{Data availability}
All data needed to evaluate the conclusions in the paper are present in the paper and/or the Supplementary Materials.

\bibliography{bib-refs}

\appendix
\section*{Appendix}

\makeatletter
\renewcommand{\thefigure}{\thesection\arabic{figure}}
\renewcommand{\thetable}{\thesection\Roman{table}}
\makeatother
\setcounter{figure}{0}
\setcounter{table}{0}

\section{Logical BSM on a two-qubit repetition code in the stabiliser framework}
\label{app:stabilisers}

The Bell states are the four simultaneous eigenstates of the two two-qubit operators $X \otimes X$ and $Z \otimes Z$. For simplicity, we will omit tensor products and keep them only where they facilitate a better understanding. Thus, an unambiguous BSM result yields the eigenvalues of these two operators. We refer to these as the $XX$- and $ZZ$-information, respectively. Since from any click pattern in Eq.~\eqref{eq:transformed-state} we are able to discriminate between the sets $\{\ket{\Phi^+},\ket{\Psi^+}\}$ and $\{\ket{\Phi^-},\ket{\Psi^-}\}$ the $XX$-information is always obtained. Since the two Bell states $\ket{\Phi^-}$ and $\ket{\Psi^-}$ always yield an unambiguous result, the $ZZ$-information is obtained if and only if the eigenvalue of $XX$ of the Bell state is $-1$.

\begin{figure}[hb!]
    \centering
    \def\svgwidth{1\linewidth}
    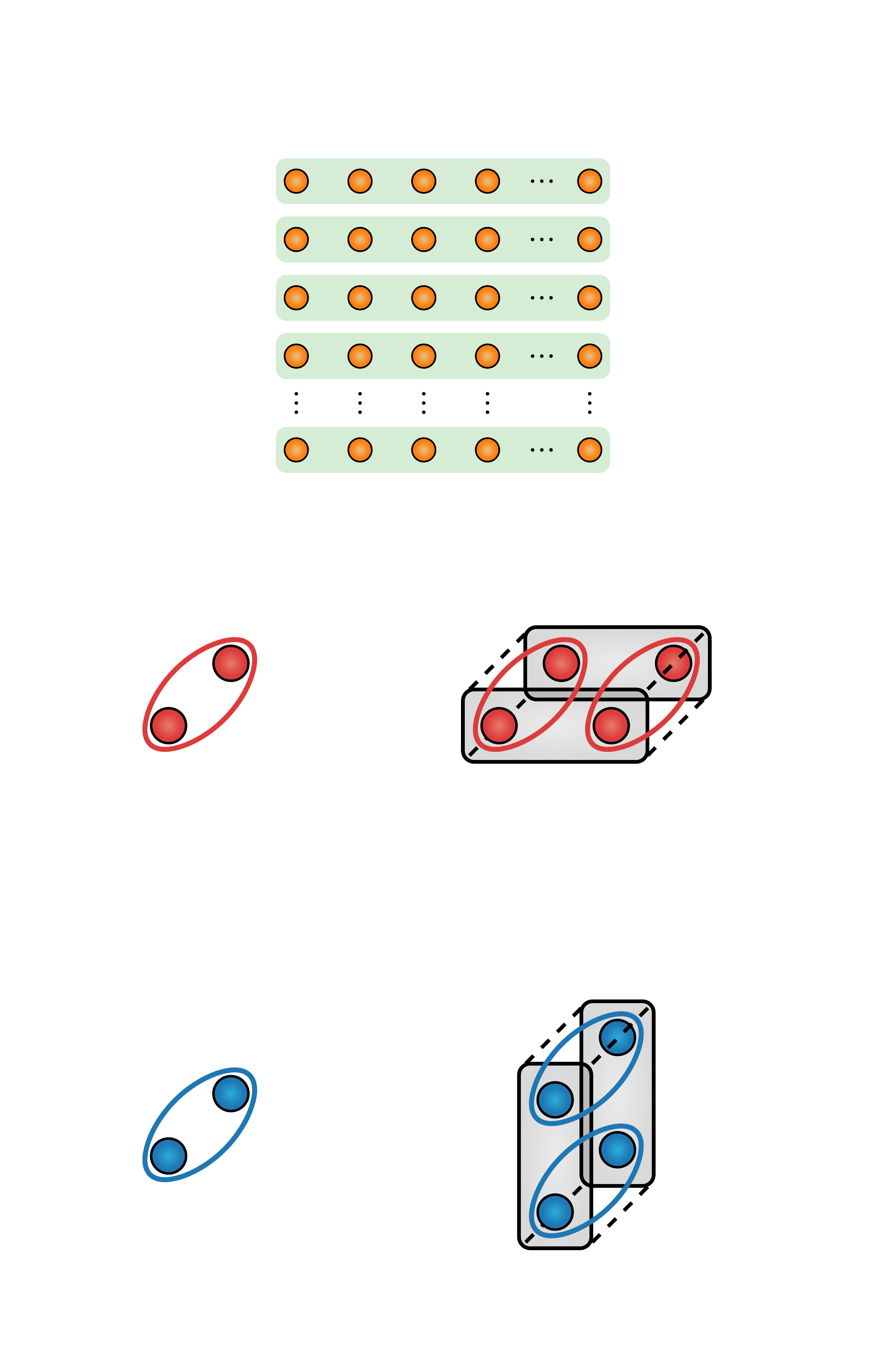
    \caption{\textbf{Logical Bell-state measurement in the stabiliser picture.} (a) QPC$(n,m)$, where the code is constructed from $n$ blocks containing $m$ qubits each. (b),(c) On the left, physical BSMs that provide guaranteed $XX$ and $ZZ$ information are depicted as red and blue ovals, respectively. On the right, schematic representations of a logical BSM on QPC($1,2$) and QPC($2,1$) are shown. Each logical qubit (grey box) is encoded in two physical qubits (red and blue circles). The physical BSMs that constitute the logical BSM are shown to act on corresponding physical qubits of the two logical qubits.}
    \label{fig:stabiliser}
\end{figure}
The QPC is a stabiliser code~\cite{gottesman1997stabilizer} and is displayed in Fig.~\ref{fig:stabiliser}(a). This code consists of $n$ blocks, each containing $m$ qubits. The structure of QPC($n,m$) naturally leads to a double-index notation, where each qubit is indexed by a pair $(i, j)$, with $i \in \{1, \dots, n\}$ denoting the block and $j \in \{1, \dots, m\}$ enumerating the qubits within each block.

Stabiliser codes are defined as the $+1$ eigenstates of a set of stabilisers. These stabilisers form a group and can therefore be specified compactly by a set of stabiliser generators. QPC($n,m$) is stabilised by two types of operators. First, each block $i$ of $m$ qubits is stabilised by $m-1$ stabiliser generators of the form $Z_{i,j}Z_{i,j+1}$ for all $j \in \{1, \dots, m-1\}$. Second, adjacent pairs of blocks $i$ and $i+1$ are stabilised by the operator $\prod_{t=1}^{m} X_{i,t} X_{i+1,t}$. In conclusion, QPC($n,m$) is defined by the stabiliser generators:
\begin{equation}
	\begin{aligned}
	G_c = & \{ Z_{i,j}Z_{i,j+1} \}_{(i,j) \in \{(i,j) \mid 1 \leq i \leq n, 1 \leq j \leq m-1 \}} \\
		& \cup \{\prod_{t=1}^{m} X_{i,t} X_{i+1,t} \}_{i \in \{1, \dots, n-1\}}.
	\end{aligned}
\end{equation}
The set $G_c$ consists of a total of $(n(m-1)+(n-1) = nm-1$ stabiliser generators, thus encoding one logical qubit in $nm$ physical ones.

The treatment of qubits on the logical level requires translating physical-level Pauli operators to logical-level Pauli operators, analogous to the translation of physical-level qubits to logical-level qubits. We denote the sets of all operators which act as Pauli $X$ and Pauli $Z$ on the logical level as $[X_L]$ and $[Z_L]$, respectively. A logical $X_L$ operator acts on each qubit of one block with $X$ operators:
\begin{equation}
	\prod_{t=1}^m X_{i,t} \in \left[ X_L \right], \quad \text{where } i \in \{ 1, \dots , n \},
	\label{eq:qpc-logical-x}
\end{equation}
and a logical $Z_L$ operator acts on one qubit in every block with a $Z$ operator:
\begin{equation}
	\prod_{t=1}^n Z_{t,j_t} \in \left[ Z_L \right],
	\label{eq:qpc-logical-z}
\end{equation}
where the indices $j_t$ can be chosen arbitrarily.

In the following, we shall focus on a special instance of the QPC, namely QPC($1,2$), as it was realised in the experiment. It is defined as the $+1$ eigenspace of the stabiliser $ZZ$, which is equivalent to the definition in Eq.~\eqref{eq:repetition-code}. The relevant logical operators of this code are:
\begin{equation}
    \{XX\} \subset [X_L], \quad \{ZI,IZ\} \subset [Z_L].
\end{equation}
Their action on the logical level can be easily verified by applying these operators to the logical basis states in Eq.~\eqref{eq:repetition-code}.
\begin{table}[ht!]
\centering
\setlength{\tabcolsep}{15pt}

\makebox[\linewidth][l]{\textbf{(a)}}
\begin{tabularx}{\linewidth}{ccc}
\toprule
&QPC($1,2$)& \\
$XX$ $1^{\rm st}$ BSM & $XX$ $2^{\rm nd}$ BSM & Logical BSM\\
\toprule
$-1$ & $-1$ & success \\
$-1$ & $+1$ & success\\
$+1$ & $-1$ & success\\
$+1$ & $+1$ & failure \\
\bottomrule
\end{tabularx}

\vspace{1em}

\makebox[\linewidth][l]{\textbf{(b)}}
\begin{tabularx}{\linewidth}{ccc}
\toprule
&QPC($2,1$)& \\
$ZZ$ $1^{\rm st}$ BSM & $ZZ$ $2^{\rm nd}$ BSM & Logical BSM\\
\toprule
$-1$ & $-1$ & success \\
$-1$ & $+1$ & success\\
$+1$ & $-1$ & success\\
$+1$ & $+1$ & failure \\
\bottomrule
\end{tabularx}

\caption{\textbf{Possible eigenvalues of the two physical BSMs.} $XX$ eigenvalues for QPC($1,2$) and $ZZ$ for QPC($2,1$) in (a) and (b), respectively. The last column shows which sets of eigenvalues lead to a successful or failed logical BSM.}
\label{tab:X_eigenvalues}
\end{table}
Combining two logical qubits, we get the logical two-qubit operators
\begin{equation}
    \begin{aligned}
        \{XX \otimes XX\} & \subset [X_L \otimes X_L], \\
        \{ZI \otimes ZI, IZ \otimes IZ\} & \subset [Z_L \otimes Z_L], \\
    \end{aligned}
\end{equation}
where the tensor product separates the physical qubits of the two codes. On the logical level, an unambiguous BSM result is achieved if two logical two-qubit operators $X_L X_L \in [X_L \otimes X_L]$ and $Z_L Z_L \in [Z_L \otimes Z_L]$ are successfully measured.

Since our physical BSMs always obtain the $XX$-information, the eigenvalue of ${XX \otimes XX} \in [X_L \otimes X_L]$ will always be obtained. For simplicity, we refer to an unambiguous BSM result as a success. If the BSM on the first qubit pair succeeds, it also obtains the eigenvalue of $ZI \otimes ZI \in [Z_L \otimes Z_L]$, and if the BSM on the second qubit pair succeeds, it also obtains the eigenvalue of $IZ \otimes IZ \in [Z_L \otimes Z_L]$. Thus, only one of the two physical BSMs needs to succeed to have a successful BSM on the logical level. If none of the physical BSMs succeeds, it is impossible to measure any element from $[X_L \otimes X_L]$ and $[Y_L \otimes Y_L]$, since any element from those sets requires either $X$ or $Y$ information on at least two qubits. A physical BSM succeeds if and only if the $XX$ eigenvalue of the input Bell state is $-1$. Therefore, of the four possible eigenvalue combinations for the Bell state, only one results in an ambiguous BSM result on the logical level, as is displayed in Tab.~\ref{fig:stabiliser}(a).

QPC($2,1$) is defined as the $+1$ eigenspace of the stabiliser $XX$. Therefore, by changing to the standard BSM without the additional Hadamard gates, which obtains the eigenvalue of $ZZ$ for all Bell states, the discussion carries over, with $X$ and $Z$ interchanged, as displayed in Fig.~\ref{fig:stabiliser}(c) and Tab.~\ref{tab:X_eigenvalues}(b). Thus, as a counterpart to QPC($1,2$), QPC($2,1$) detects random phase-flip errors rather than bit-flip errors.

\section{Resource state generation and source characterisation}
\label{GHZ_Appx}

We generate the logical states using two Bell pairs as resource states and interfering one photon from each pair using a PBS. These Bell pairs are generated using a Sagnac-type SPDC source. We characterise the generated states by measuring their visibilities, defined by 

\begin{equation}
    V=\frac{N_{\text{desired}}-N_{\text{undesired}}}{N_{\text{desired}}+N_{\text{undesired}}},
\end{equation}
in the $ Z$- and $ X$-bases. Here, $N_{\text{(un)desired}}$ refers to the coincidence counts of the (un)desired measurement outcomes. We estimate a visibility of $(99.02\pm0.08)\%$ and $(99.46\pm0.04)\%$ in the $Z$-basis and $(98.09\pm0.11)\%$ and $(99.33\pm0.05)\%$ in the $X$-basis for the two sources, respectively. 

Since visibility has an upper bound of $100\%$, a symmetric Gaussian error can result in an upper bound exceeding this physical limit. To account for this, we truncate the distribution at $100\%$ and estimate the corresponding confidence interval using the truncated cumulative distribution function (CDF), which can sometimes result in asymmetric errors. 

Here, the visibilities remain unchanged after truncation, and the errors are estimated assuming a Poissonian distribution for the photon-detection process.

\section{Two-source interference using PBS}
\begin{figure}[t!]
    \centering
    \includegraphics[width=1\linewidth]{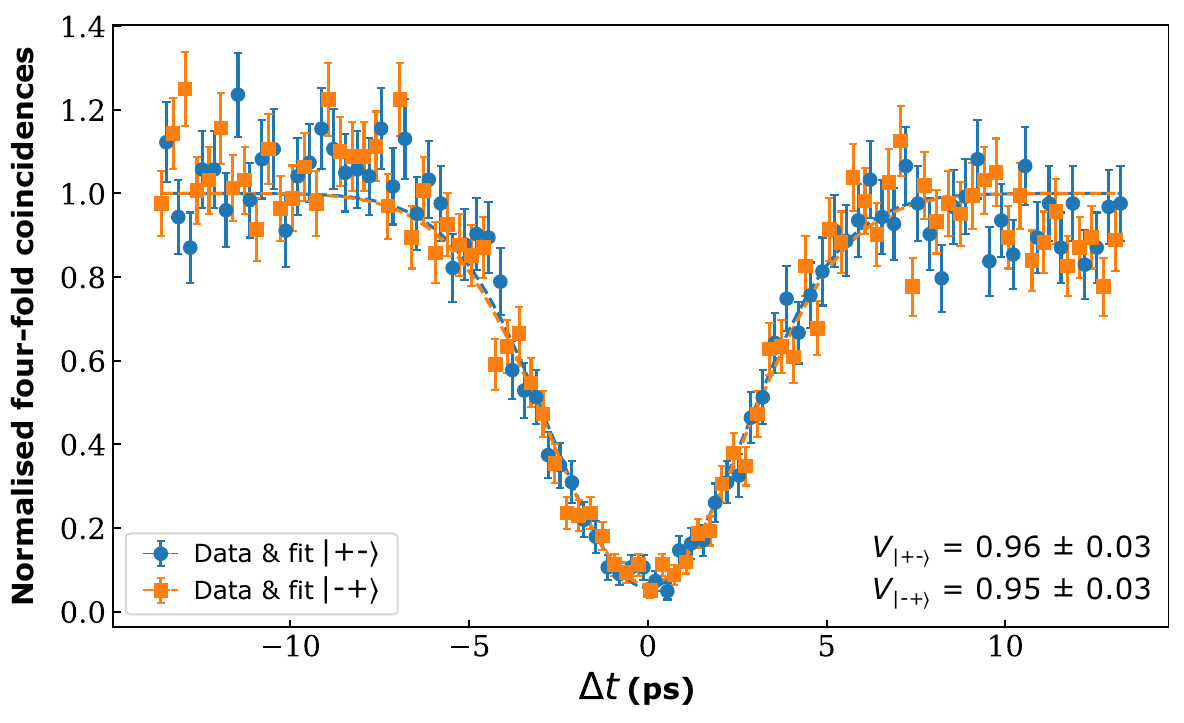}
    \caption{\textbf{Interference between the two photon-pair sources.} Heralded two-source interference between photons from the two Sagnac sources measured at a PBS. The experimental data are fitted using a Gaussian function, from which the interference visibilities are extracted.}
    \label{fig:Two-sourcePBS}
\end{figure}
We measure the interference of the photons from the two sources by performing heralded interference with one photon from each source at the PBS. We introduce polarisers at both inputs of the PBS, oriented along the diagonal polarisation setting. The input state is therefore 
\begin{equation}
\begin{split}
      \ket{\psi_{\text{in}}}=&a^{\dagger}_+ b^{\dagger}_+\ket{\text{vac}}\\
      =&\frac{1}{2}\left(a^{\dagger}_H b^{\dagger}_H+a^{\dagger}_H b^{\dagger}_V+a^{\dagger}_V b^{\dagger}_H+a^{\dagger}_V b^{\dagger}_V\right)\ket{\text{vac}},   
\end{split}
\end{equation}

where $a$ and $b$ are the two modes of the PBS. The PBS reflects vertically polarised light while transmitting horizontally polarised light. This results in
\begin{equation}
    \ket{\psi_{\text{out}}}=\frac{1}{2}\left(a^{\dagger}_H b^{\dagger}_H+a^{\dagger}_H a^{\dagger}_V+b^{\dagger}_V b^{\dagger}_H+b^{\dagger}_V a^{\dagger}_V\right)\ket{\text{vac}}.
\end{equation}
Post-selecting on coincidences at the output results in 
\begin{equation}
    \ket{\psi_{\text{post}}}=\frac{1}{\sqrt{2}}\left(a^{\dagger}_H b^{\dagger}_H+a^{\dagger}_V b^{\dagger}_V\right)\ket{\text{vac}},
\end{equation}
with a probability of 50\%. In the $X$-basis, this transforms to
\begin{equation}
    \ket{\psi_{\text{post}}}=
    \frac{1}{\sqrt{2}}\left(a^{\dagger}_+ b^{\dagger}_++a^{\dagger}_- b^{\dagger}_-\right)\ket{\text{vac}},
\end{equation}
if the two photons are indistinguishable. This results in a dip in the coincidences for $a^{\dagger}_+ b^{\dagger}_-\ket{\text{vac}}$ and $a^{\dagger}_- b^{\dagger}_+\ket{\text{vac}}$. Experimentally, we measure visibilities of $(96\pm3)\%$ and $(95\pm3)\%$ for the two polarisation combinations, respectively, as presented in Fig.~\ref{fig:Two-sourcePBS}. Here, the visibilities are truncated to $(94\pm3)\%$ and $(95^{+2}_{-3})\%$. 

The reduction in visibility can be attributed to the imperfect extinction ratio of the bulk PBS. For comparison, the maximum visibility measured at the PBS with a single source was $(96.9\pm0.4)\%$. Truncation does not lead to a change in this case. The remaining difference between the single-source and two-source visibilities could arise from residual mode mismatch, spectral impurity of the interfering photons, and higher-order pair generation in the SPDC source. We estimated the heralded second order correlation ($g^{(2)}(0)$) of the two sources to be $0.0224\pm0.0019$ and $0.0163 \pm 0.0013$, respectively. This corresponds to an average probability of higher orders of about 0.5\%.

\section{Interference at $50:50$ beam splitters}
\setcounter{figure}{0}
\begin{figure}[t!]
    \centering
    \includegraphics[width=1\linewidth]{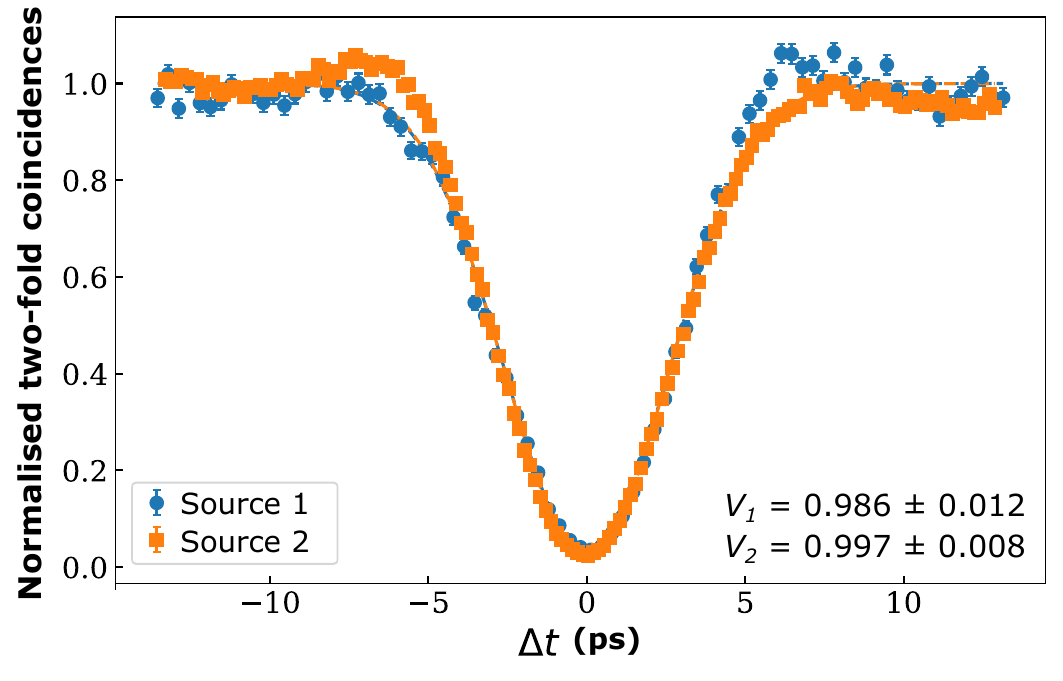}
    \caption{\textbf{Single-source HOM interference at a balanced beam splitter.} HOM interference measurements between photon pairs from the same source to ensure temporal indistinguishability at the balanced beam splitters.}
    \label{fig:single_source_dips}
\end{figure}
The logical BSM is implemented by performing two standard BSMs. We realise this experimentally using two fibre-based balanced beam splitters. To ensure temporal indistinguishability of the photons, we perform single-source HOM interference using the two beam splitters.
The resulting characteristic dips are plotted in Fig.~\ref{fig:single_source_dips}. From the Gaussian fits, we estimate a visibility of $(98.6\pm1.2)\%$ for the first source at the upper beam splitter, and $(99.7\pm0.8)\%$ for the second source at the lower beam splitter (see Fig.~\ref{fig:experimental_implementaion}), which are truncated, as described previously, to $(98.42^{+1.1}_{-0.9})\%$ and $(99.35^{+0.6}_{-0.4})\%$, respectively. This indicates a high degree of indistinguishability of the single photons from the individual sources.

\section{Pseudo-Photon-Number Resolution}
Our scheme requires photon-number resolution up to two photons. While there have been recent developments on photon-number resolving detectors, we employ pseudo-PNR by spatially demultiplexing each of the output channels of the standard BSMs to four click detector channels. The idea is that if multiple photons arrive in the original mode, they are probabilistically distributed among these four sub‑modes, and the pattern of detector clicks can be used to infer how many photons were present. The probability that $n$ photons are resolved by $k$ detector channels is given by ~\cite{bayerbach2023,hauser2025}
\begin{equation}
    P(n,k)=\frac{k!}{(k-n)!k^n}.
\end{equation}
In our case, we resolve 2 photons using 4 modes, resulting in a factor of 0.75. The measured rate of detecting $n$ photons in $k$ modes is $P(n,k)$ times the true rate in the demultiplexed mode. Consequently, each output mode is rescaled by a correction factor of $1/P(n,k)$. 

\section{Background correction and phase randomisation}
We use probabilistic SPDC sources that exhibit a finite probability of generating multiple photon pairs simultaneously. Without heralding any of the photons, there can be four-fold coincidence events that do not originate from the intended logical Bell state, but instead from higher-order contributions. To mitigate this effect, we perform a background correction by subtracting four-fold background counts. These background counts are obtained in separate measurements of equal duration, during which one source at a time is blocked, thereby isolating and quantifying the higher-order noise contribution.

Furthermore, the coincidence probability acquires an interference term that depends on the relative phase between the two sources. We randomise this phase by introducing a QWP--HWP--QWP stack with both QWPs oriented at $45^\circ$, whilst continuously rotating the HWP during the measurement. 
\end{document}